# The INFN-FBK "Phase-2" R&D Program

G.-F. Dalla Betta[a,b,*], M. Boscardin[c,b], M. Bomben[d], M. Brianzi[e], G. Calderini[d,f,g], G. Darbo[h], R. Dell'Orso[g], A. Gaudiello[i,h], G. Giacomini[c], R. Mendicino[a,b], M. Meschini[e], A. Messineo[f,g], S. Ronchin[c], D M S Sultan[a,b], N. Zorzi[c,b]

[a] *Università di Trento, Dipartimento di Ingegneria Industriale, I-38123 Trento, Italy*
[b] *TIFPA INFN, I-38123 Trento, Italy*
[c] *Fondazione Bruno Kessler (FBK), I-38123 Trento, Italy*
[d] *Laboratoire de Physique Nucleaire et de Hautes Énergies (LPNHE), 75252 Paris, France*
[e] *INFN Sezione di Firenze, I-50019 Sesto Fiorentino, Italy*
[f] *Università di Pisa, Dipartimento di Fisica, I-56127 Pisa, Italy*
[g] *INFN Sezione di Pisa, I-56127 Pisa, Italy*
[h] *INFN Sezione di Genova, I-16146 Genova, Italy*
[i] *Università di Genova, Dipartimento di Fisica, I-16146 Genova, Italy*

**Abstract**

We report on the 3-year INFN ATLAS-CMS joint research activity in collaboration with FBK, started in 2014, and aimed at the development of new thin pixel detectors for the High Luminosity LHC Phase-2 upgrades. The program is concerned with both 3D and planar active-edge pixel sensors to be made on 6" p-type wafers. The technology and the design will be optimized and qualified for extreme radiation hardness ($2\times10^{16}$ $n_{eq}$ cm$^{-2}$). Pixel layouts compatible with present (for testing) and future (RD53 65nm) front-end chips of ATLAS and CMS are considered. The paper covers the main aspects of the research program, from the sensor design and fabrication technology, to the results of initial tests performed on the first prototypes.

*Keywords*: High Luminosity LHC; 3D silicon sensors; Active Edges; Fabrication technology.

## 1. Introduction

The upgrades at the High Luminosity LHC (HL-LHC) will need the complete replacement of the ATLAS and CMS inner trackers with new ones fulfilling the requirements of higher radiation fluences ($2\times10^{16}$ $n_{eq}$ cm$^{-2}$, or equivalently 1 Grad, expected on the inner pixel layer for 2500 fb$^{-1}$ integrated luminosity in the Phase-2), and higher event pile-up (140 events/bunch-crossing) [1]. To maintain the same performance of the present detector systems a new generation of technologies has to be fully exploited for the redesigned Pixel detectors. Among them the future version of front-end chips in 65-nm CMOS by the CERN RD53 Collaboration will allow for smaller pixel sizes (50×50 or 25×100 μm$^2$) and lower thresholds (~1000 e) [2]. The advances in the front-end design shall require sensors with smaller pixel cells and thinner active thickness to match the reduced pixel dimension and to improve track resolution and cluster separation in higher pile-up environment. Additional optimization of the new Pixel detector requires the reduction of the radiation-length of the layer to minimize secondary interactions and Multiple Coulomb Scattering effects.

To this purpose, a new generation of 3D sensors [3] and of planar sensors with active edges (PAE) [4]

---

are being developed in the framework of the INFN Phase-2 program, and will be fabricated at the pilot line of FBK (Trento, Italy), that was recently updated to 6-inch wafers. Another partnership is in place with Selex SI (Rome, Italy) for further developing Indium bump-bonding technology, that is potentially more suited than solder-reflow for large chip sizes, thinner electronic and sensor substrates due to the lower temperature of the process (90 ºC instead of 250 ºC).

This paper will focus on the main aspects of the research program, with emphasis on the sensor design and fabrication technology, and will report initial results from the electrical tests on the first prototypes.

## 2. Technological aspects

Increased luminosity requires higher hit-rate capability, increased granularity, higher radiation tolerance, and reduced material budget. Most of these requirements benefit from having thinner active layers. Among the possible substrate options suitable for the fabrication of thin pixels (e.g., SOI, epitaxial, or local thinning [5]), we have chosen to fabricate pixel detectors on Si-Si Direct Wafer Bonded (DWB) wafers, which are obtained bonding together two different wafers: a high-resistivity (HR) Float Zone sensor wafer and a low-resistivity (LR) Czochralski handle wafer. The FZ wafer is thinned to the desired thickness value, so as to obtain a wafer with a thin active layer plus a relatively thick mechanical support layer. P-type wafers of two different active depths (100 and 130μm) with 500-μm thick handle wafer were purchased from IceMOS Technology Ltd. (Belfast, UK).

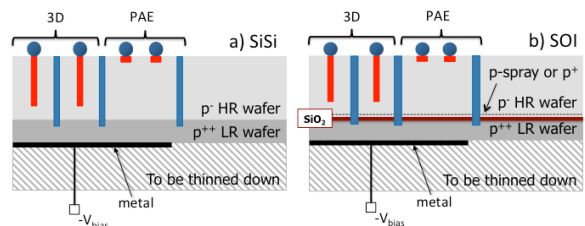

Figure 1 Schematic cross-section of the proposed thin 3D and planar active-edge sensors on (a) SiSi DWB substrate, and (b) SOI substrate.

A schematic cross section of the technological approach being developed for 3D and PAE sensors on these substrates is shown in Fig. 1a. For 3D sensors, column etching is performed by Deep Reactive Ion Etching (DRIE) from the front-side of the wafer with two different depths: slightly shorter (~15 μm) than the active layer for junction ($n^+$)
columns, so as to obtain a high breakdown voltage; slightly deeper than the active layer for ohmic ($p^+$) columns, so that a good ohmic contact is obtained by the highly doped handle wafer, making sensor bias possible from the back side. To this purpose, as a post processing step combined with the bump bonding process, the handle wafer will be thinned and a metal layer will be deposited on the back side. A special technology has been developed at FBK for these new 3D sensors, which differ significantly from those produced for the ATLAS Insertable B-Layer [6] both in terms of electrode dimensions and of process details, as reported in [7].

Similarly, for PAE sensors (Fig. 1a), the junction ($n^+$) electrodes are confined at the front surface, whereas deep trenches are etched to reach the handle wafer and doped to act as ohmic ($p^+$) contacts.

As an alternative, we are also considering SOI wafers, which differ from the Si-Si DWB ones for the presence of a 200-nm thick $SiO_2$ layer in between the two silicon layers. In order to adopt the same technological approach, in this case it is necessary to etch columns (3D) or trenches (PAE) through the bonding $SiO_2$ layer, so as to reach the LR handle wafer. This step is not trivial, and dedicated tests have been performed at FBK to prove its feasibility. As an example, Fig. 2 shows the sketch of the performed test, which consisted in etching columns all the way through a 200-μm thick wafer coated with a 200-nm thick $SiO_2$ layer on both sides, so as to mimic the final structures. A 2-μm thick poly-Si layer is present on the wafer back side as an etch stop. The SEM micrograph in Fig. 2 demonstrates that such an etching is indeed feasible, yet the process should be optimized to improve uniformity and reproducibility before using it in a real production.

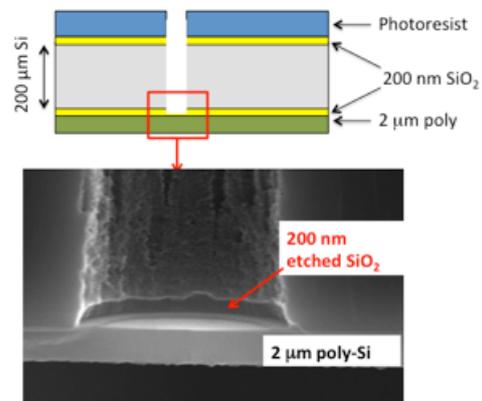

Figure 2 Sketch of the process test made to investigate deep etching through the bonding oxide layer, and SEM micrograph of the bottom of a column etched by DRIE down to the poly-Si layer.

## 3. Results from first prototypes

A first batch of sensors was processed at FBK in 2014 with a n-on-p planar technology using a wafer layout mainly based on ATLAS FEI4 and CMS PSI46 designs. Standard and specific test structures were also included with a high multiplicity allowing for statistically significant results. The goal of this batch was to evaluate the properties of SiSi DWB wafers, here used for the first time, and to start testing thin n-on-p planar pixel sensors, also including some issues relevant to their high voltage operation (e.g., spark protection by benzo-cyclo-butene [8]).

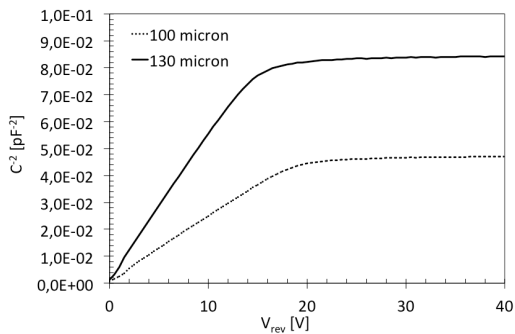

Figure 3 $1/C^2$-V curves measured on two test diodes from SiSi DWB wafers of different active thickness.

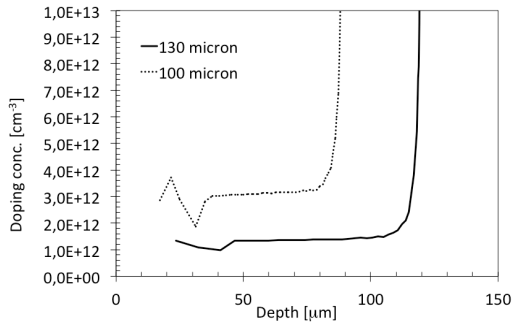

Figure 4 Doping concentration as a function of depth as extracted from the $1/C^2$-V curves of two test diodes from SiSi DWB wafers of different active thickness. Depletion caused by the built-in voltage prevents from extracting meaningful values at small depth

From the electrical characterization of test structures, measured on wafer with a probe station, it was possible to assess the quality of both the raw material and of the fabrication process. Fig. 3 shows the capacitance as a function of reverse bias in two diodes with guard ring from wafers of different active thickness: $1/C^2$-V curves are shown to better appreciate the full depletion voltage, that is lower than 20 V in both cases. Notably, the depletion voltage is lower for the thicker device, evidence of a different doping concentration. This is confirmed by Fig. 4, which shows the doping concentration profiles extracted from the $1/C^2$-V curves: within the depth intervals where they are meaningful, doping concentrations are indeed different by about a factor of 3. Both profiles start deviating from a constant concentration as the depth approaches the bottom of the active layer. In both cases, this happens about 10 μm below the nominal thickness, as a result of two concurrent factors: the back diffusion of boron from the highly doped handle-wafer (estimated by IceMOS to be about 5 μm) and a C-V measurement effect arising from the Debye length within the HR layer.

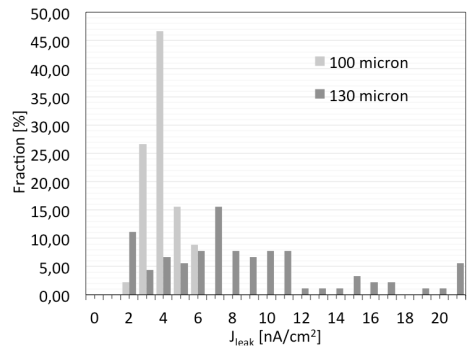

Figure 5 Leakage current density distribution at full depletion for 135 test diodes from SiSi DWB wafers of different active thickness.

The leakage current was measured on all test diodes showing good values. As an example, Fig. 5 shows the distribution of the leakage current density at full depletion in a sample of 135 test diodes from wafers of different active thickness. All diodes exhibit values from 2 nA/cm$^2$ to about 20 nA/cm$^2$. The distribution is peaked at low values for the 100-μm thick devices, whereas it is broader for the 130-μm thick ones. This difference, and the observed difference in the doping concentration, is most likely due to different ingots of FZ material used in these SiSi DWB substrates.

From I-V curves of test diodes (not shown) it was also possible to measure breakdown voltages in the range from ~120 V to ~160 V, properly scaling with the three different p-spray doses used in different wafers. Pixel sensors were also electrically tested, confirming the good results in terms of leakage current and actually exhibiting much larger breakdown voltages, from ~160 V to ~500 V, owing to the use of multiple guard rings. As an example, Fig. 6 shows the I-V curves for a set of CMS pixel sensor with different guard ring terminations.

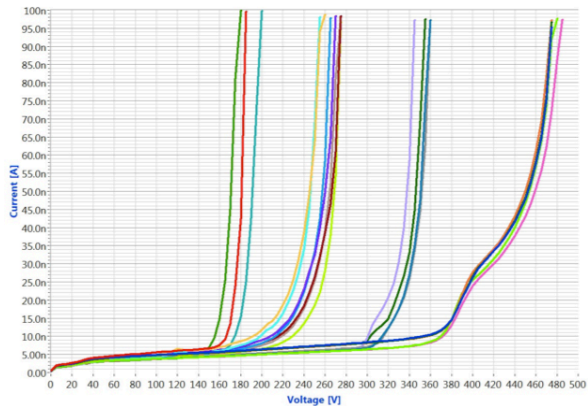

Figure 6 I-V curves of CMS pixel sensors with different guard ring terminations.

The best five wafers of this batch, having an electrical yield higher than 80% on pixel sensors, are being bump-bonded at IZM (Berlin, Germany) and at Selex. The design of a batch of PAE sensors is under way: different guard ring solutions, aimed at maximizing the breakdown voltage while limiting the size of the edge region, will be explored [9,10].

The fabrication of a first batch of 3D sensors has just been started at FBK [7]. The design includes several variants of pixel sensors, allowing to investigate small pixel dimensions (i.e., $50 \times 50$ μm$^2$ and $25 \times 100$ μm$^2$), while ensuring compatibility with the existing read-out chips (ROCs): ATLAS FE-I4 (with $50 \times 250$ μm$^2$ native pixels) and CMS PSI46 ($150 \times 100$ μm$^2$).

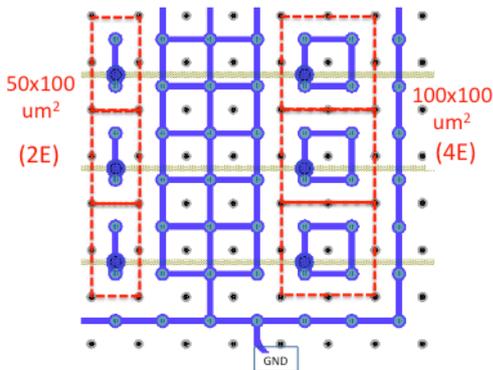

Figure 7 Layout detail of a 3D pixel sensor featuring $50 \times 50$ μm$^2$ cell sizes while being compatible with CMS PSI46 read-out chip.

To this purpose the sensor layouts place n and p columns on either 25 μm ×100 μm or 50 μm × 50 μm grids that define elementary cells. One or more cells are then connected to the bonding pads of the ROC, while the remaining n columns all shorted by a metal grid connected to extra bonding-pads that are grounded in the ROC. As an example, Fig. 7 shows a layout compatible with a CMS PSI46 ROC and featuring 2E and 4E pixel configurations alternated to pixels grounded by grid. Other pixel sensors are present compatible with the first RD53 prototypes.

Although the fabrication technology allows for active edges, in the first batch we preferred to use slim edges, featuring either 3D guard rings or the ohmic column fence introduced in [11]. Owing to the higher density of narrow columns now adopted [7], the slim edge design can be optimized, leading to a dead area at the edge lower than 100 μm, that is good enough for the intended application.

## 4. Conclusion

We have reported on the INFN-FBK R&D program aimed at new thin pixel sensors for HL-LHC. SiSi DWB wafers have been explored for the first time with electrical figures, functional tests are under way. With reference to both SiSi DWB and SOI wafers, new design solutions and fabrication processes have been developed for 3D and PAE sensors. The first 3D batch is being fabricated at FBK, while the design of the PAE batch is being completed and fabrication at FBK will start soon.

**Acknowledgment**

This work was supported by the Italian National Institute for Nuclear Physics (INFN), Projects ATLAS, CMS, RD-FASE2 (CSN1), and by AIDA-2020 project EU-INFRA proposal no. 654168.

**References**

[1]  F. Gianotti et al., Eur. Phys. J. C 39 (2005) 293.
[2]  M. Garcia-Sciveres, J. Christainsen, CERN-LHCC-2013-002, LHCC-I-024. http://rd53.web.cern.ch/RD53
[3]  C. Da Via et al., Nuclear Instruments and Methods in Physics Research Section A 694 (2012) 321.
[4]  C.J. Kenney et al., Nuclear Instruments and Methods in Physics Research Section A 565 (2006) 272.
[5]  G.-F. Dalla Betta, PoS(IFD2014)013.
[6]  G. Giacomini et al., IEEE Transactions on Nuclear Science NS-60 (3) (2013) 2357.
[7]  G.-F. Dalla Betta et al., "Development of a new generation of 3D pixel sensors for HL-LHC", These Proceedings.
[8]  C. Gallrapp et al., Nuclear Instruments and Methods in Physics Research Section A 679 (2012) 29.
[9]  M. Povoli et al., Nuclear Instruments and Methods in Physics Research Section A 658 (2011) 103.
[10] M. Bomben et al., Nuclear Instruments and Methods in Physics Research Section A 730 (2013) 215.
[11] M. Povoli et al., J. of Instrumentation 7 (2012) C01015.